\documentclass[pra,aps,epsfig,amsmath,amssymb,showpacs]{revtex4}

\newcommand{\re}{{\rm e}}
\newcommand{\ri}{{\rm i}}

\newcommand\be{\begin{eqnarray}}
\newcommand\ee{\end{eqnarray}}
\newcommand\nn{\nonumber}
\newcommand{\abs}[1]{\vert#1\vert}

\usepackage{graphicx}

\begin{document}
\bibliographystyle{prsty}
\title{Relation between discrete and continuous 
teleportation using linear elements}
\author{Dirk Witthaut and Michael Fleischhauer}
\affiliation{Fachbereich Physik, 
Universit\"{a}t Kaiserslautern, D-67663 Kaiserslautern,
Germany}
\date{\today}

\begin{abstract}
We discuss the relation between discrete and continuous linear teleportation, 
i.e. teleportation schemes that use only linear optical elements and 
photodetectors. For this the existing
qubit protocols are generalized to qudits with a discrete and
finite spectrum but with an arbitrary number of states or alternatively 
to continuous variables. Correspondingly a generalization of 
linear optical operations
and detection is made. It is shown that linear teleportation is 
only possible in a probabilistic sense.
A general expression for the 
success probability is derived which is shown to depend only on the dimensions of
the input and ancilla Hilbert spaces. From this the known results $P=1/2$ and
$P=1$ for the discrete and continuous cases can be recovered. 
We also discuss the probabilistic teleportation scheme of Knill, Laflame and
Milburn and argue that it does not make optimum use of resources.
\end{abstract}


\maketitle


\section{Introduction}


Recently Knill, Laflamme and Milburn (KLM) \cite{Knill-Nature-2001} proposed a scheme 
for linear optical quantum computing using photons as qubits. 
This scheme works efficient if one can implement deterministic 
discrete quantum teleportation. 
The original proposal  for discrete teleportation by
Bennett {\it et.al.} \cite{Bennett-PRL-1993} involves the projection 
on two-particle Bell states which, as shown in 
\cite{Lutkenhaus-PRA-1999,Calsamiglia-PRA-2002}, cannot
be implemented with linear elements in a deterministic way, unless
entanglement in additional degrees of freedom is used 
\cite{Cerf-PRA-1998,Kwiat-PRA-1998}.
{\it Discrete} teleportation with linear elements 
is only possible using post selection
with a success probability of 50 \% e.g. using the scheme suggested by
Weinfurter \cite{Weinfurter-EPL-1994}.
On the other hand in the case of {\it continuous} variables,
perfect teleportation with linear elements can be achieved
as shown by Braunstein and Kimble \cite{Braunstein-PRL-1998}.
This raises the
question about the origin of this manifestly different behavior. 
Vaidman and Yoran suggested that the beam-splitter used in the
Braunstein-Kimble experiment would lead to an effective ``quantum-quantum''
interaction \cite{Vaidman-PRA-1999,Vaidman-PRA-1994}. The proposal of KLM hints
in a different direction: The success probability
of the KLM teleportation protocol tends toward 100 \% when an increasing 
number of additional ancilla photons is used \cite{Knill-Nature-2001},
which indicates that the difference could be connected to the
resources used.

We here discuss a general relation between discrete and continuous
teleportation using linear elements. We show that the 
success probability of a generalized linear  protocol 
is only limited by the
dimensions of the input and ancilla spaces. It is argued that the
KLM scheme is inefficient from this point of view as it does not
reach this limit. Recently Franson et.al. \cite{Franson-PRL-2002} suggested a 
modification of the KLM 
scheme with a better scaling of the fidelity with the number
of ancilla photons at the expense that the
input state is never reconstructed exactly. This scheme also does not reach
the limit obtained in the present paper. 

Guided by continuous variable teleportation protocols 
\cite{Braunstein-PRL-1998,Vaidman-PRA-1994} we 
generalize teleportation from qubits to qudits with
a bounded discrete spectrum of states. 
Also the notion of linear elements is generalized 
reducing however to photodetection and / or homodyne
detection in the two limits.
We will restrict the discussion first to schemes 
that faithfully reproduce the input state, i.e. in which Alice and Bob
can decide whether the teleportation was successful. 
At the end we will also discuss the connection to teleportation schemes of the 
Franson-type with non-unity fidelity but higher success probability.


\section{A generalized linear teleportation protocol}


Let us briefly recall the general properties of a teleportation protocol:
Alice possesses an unknown quantum state $\left| \Psi \right\rangle$ 
in a Hilbert space of  dimension $d$ (qubits for $d=2$
or qudits in the general case),
which she wants to transfer to Bob by only applying local operations including 
measurements and exchanging classical information. 
In the following we will restrict ourselves to discrete generalizations
of a qubit, note however that also continuous generalizations can be
discussed in a similar way. 
For the teleportation Alice and Bob make use
of an ancilla system with state 
$\left| \Phi \right\rangle$, which is
an entangled pair of qudits which they share.
Alice then performs an appropriate measurement on her qudit of
the ancilla and the unknown quantum state. This measurement projects Bob's
ancilla qudit to a certain state which is up to local operations identical to
the unknown one. Alice's measurement must be such that the knowledge about its
outcome, transmitted via classical channels, is
sufficient for Bob to reconstruct the initial state $|\Psi\rangle$ 
by local operations.
If Alice is able to perform all possible
measurements in the joint Hilbert space of her qudit of the ancilla state  
and the unknown state,
unconditional teleportation is always possible. Unfortunately
this is in general a very difficult task. For example in order to 
measure the four Bell states in the Hilbert space of two qubits, as required 
in the teleportation protocol of Bennett {\it et.al.}, quantum gate operations
must be performed. As shown in \cite{Lutkenhaus-PRA-1999,Calsamiglia-PRA-2002}
for the case of photonic qubits, only
two out of the four Bell states can be distinguished without gate 
operations by linear optical elements and photodetection.
On the other hand in the case of continuous teleportation linear elements 
suffice. Until now the limitations of teleportation with linear elements are
not completely understood. Since linear elements are easy to implement such an
understanding is very important for the practical realization
of quantum information processing. 
We thus restrict the discussion
in the following to teleportation protocols that use only 
certain generalizations of linear optical
operations and photodetection. 

Let us consider an unknown input state decomposed
into a set of orthonormal basis 
states $\left| q \right\rangle$, which are (non-degenerate) eigenvectors 
of an observable $\hat q$
\be
\hat q \, |q\rangle =q\, |q\rangle.
\ee
The real numbers $q$ are the eigenvalues
which are assumed to lie in the symmetric interval $\{-a,a\}$ with integer
steps. Their total number is $2a+1$.
Thus Alice's state is given by 
\be
|\Psi\rangle &=& \sum_{q_1 = -a}^{a} \alpha_{q_1} \left| q_1 \right\rangle_A.
\ee
The ancilla state shared by Alice and Bob has the general form:
\be
|\Phi\rangle &=& 
\sum_{q_2,q_3 = -b}^b \beta_{q_2q_3} \left| q_2 \right\rangle_A \left|
  q_3 \right\rangle_B, 
\ee
where we have assumed for the sake of simplicity that $q_2$ and $q_3$ 
have the same symmetric interval of allowed values $\{-b,b\}$, again with
integer steps.
In order to teleport an unknown state from a Hilbert space of dimension
$2a+1$, there must be at least the same number of 
pairs of bi-orthogonal ancilla
states, i.e.
the Schmidt number of the ancilla state must be larger or equal
to $2a+1$, implying $b \ge a$. 
Without loss of generality we here require that the 
coefficients $\beta$ are such that
only states that fulfill the relation $q_3=-q_2$ have a non-vanishing
amplitude, and that the Schmidt number is $b$.
This choice is quite general since any state in which 
there is a unique relation between
the eigenvalues $q_2$ and $q_3$ can be brought into this form by reordering.
One further needs that
\[
  \abs{\beta_{q_2}} = \textrm{const.} \quad \forall q_2.
\]
This condition can be deduced 
in a general way for tight teleportation schemes \cite{Werner-JPA-2001},
but will also become clear in the course of the present discussion.
 For
simplicity we choose all $\beta_{q_2}$ to be equal, i.e.
\be
  |\Phi\rangle &=& \frac{1}{\sqrt{2b+1}}  \sum_{q_2= -b}^b 
  \left| q_2 \right\rangle_A \left| -q_2 \right\rangle_B.
\ee
The above scheme includes in particular the input and ancilla states of the
Bennett protocol
as well as the KLM protocol.  In the latter case the 
state $\left| q_i \right\rangle$ 
is a quantum state of $n$ modes occupied by $q_i$ photons.

For the teleportation Alice needs to measure the total state 
$|\Gamma_0\rangle= |\Phi\rangle \otimes |\Psi\rangle$  
in a way that the input state is restored in Bob's qudit
up to a local operation which is uniquely defined by the
outcome of the measurement. As discussed above, we will 
not allow for all
possible measurements but restrict ourselves to generalized 
{\it linear} measurements, which will be defined in the following.

In an optical realization the 
$\hat q_i$ correspond to either photon numbers
or quadrature amplitudes of an electromagnetic field 
which can be measured by direct photon counting
or homodyne detection. A measurement scheme that uses
only linear optical elements like beamsplitters etc. can only
project onto eigenstates of linear combinations of photon number
or quadrature amplitude operators. 
We thus call a generalized linear measurement a
projection onto eigenstates of any linear combination 
of operators $\hat x_1$ and $\hat x_2$, i.e. of operators 
$\hat X$ in ${\cal H}_1\oplus {\cal H}_2$.

Without loss of generality we consider here a 
projection onto eigenstates of 
\be
\hat Q_+=\hat q_1 +\hat q_2,
\ee
which corresponds to the setup used in continuous teleportation.
Measuring $\hat Q_+$ can lead to only $2a+2b+1$ different
outcomes. The total Hilbert space ${\cal H}_1\otimes {\cal H}_2$
is however of dimension $(2a+1)\times(2b+1)$ and thus the spectrum
of eigenstates of $\hat Q_+$ must be degenerate. In order to
project onto a non-degenerate state a second measurement is required.
In the continuous teleportation protocol of Braunstein and Kimble
\cite{Braunstein-PRL-1998}, where $\hat q_{1,2}$ are equivalent to
position operators, the difference between the two momenta $\hat P_-=\hat p_1
-\hat p_2$ is measured in addition to $\hat Q_+$. \\
We now define an analogue measurement in the discrete case. To this end we 
construct another orthonormal basis $| p \rangle$ with the property
\be
  \abs{\langle p | q \rangle} = \text{const.} \quad \forall \, p,q.
\ee
We define the states
\begin{eqnarray}
  | p \rangle &:=&  \frac{1}{\sqrt{2b+1}} \sum_{q = -b}^b \exp\left\{
{\frac{2\pi \ri 
  qp}{2b+1}}\right\}\, | q \rangle \label{eqn-transformation_pq}
\end{eqnarray}
with $p$ running in integer steps from $-b$ to $b$. Furthermore we assume that 
the quantum number $p$ can be detected by measuring  the observable
\be
  \hat p = \sum_{p = -b}^b p |p\rangle\langle p |.
\ee
Again we allow only measurements of linear combinations of
the operators $\hat p_1$ and $\hat p_2$, i.e.
linear measurements. 
In particular Alice could measure 
 in addition to $\hat Q_+$ the quantity
\be
  \hat P_- = \hat p_1 - \hat p_2,
\ee
in close analogy to continuous teleportation.
The measurement operators $\hat Q_+$ and $\hat P_-$ do not commute, except
in a limiting case, but
this does not matter as we will see. For simplicity 
of the discussion we
consider a measurement outcome with $Q \ge 0$ and $P \ge 0$. Then the
measurement outcome is described by the projection operators:
\begin{eqnarray}
  \hat \Pi_Q &=& \sum_{q = Q-b}^b |Q-q,q \rangle \langle Q-q,q| \\
  \hat \Pi_P &=& \sum_{p=-b}^{b-P} |P+p,p \rangle \langle P+p,p|.
\end{eqnarray}
The subsequent measurement of $\hat Q_+$ and $\hat P_-$ thus projects the initial
state $| \Gamma_0 \rangle$ onto
\be
  | \Gamma_1 \rangle = \frac{\hat \Pi_P \hat \Pi_Q | \Gamma_0 \rangle}{
  \| \hat \Pi_P \hat \Pi_Q | \Gamma_0 \rangle \| }.
\ee
To calculate this expression
it is convenient to express the  projection
operators in one common basis, so we express $\hat \Pi_P$ in terms of 
the states $| q \rangle$ via eqn. (\ref{eqn-transformation_pq}):
\be
  \hat \Pi_P &=&  \frac{1}{2b+1} \sum_{p = -b}^{b-P} \,\, 
\sum_{q', \bar q' = -b}^b
  \,\,\sum_{q'',\bar q'' = -b}^b  |q',q''\rangle \langle \bar q',\bar q''| \nn \\
  && \qquad \times \exp\left\{ \frac{2\pi \ri}{2b+1} \bigl( 
  q'(P+p) + q''p - \bar q'(P+p) - \bar q'' p\bigr) \right\}.
\ee
The total projection operator is thus given by
\begin{eqnarray}
  \hat \Pi_P \hat \Pi_Q &=& \frac{1}{2b+1} \sum_{p = -b}^{b-P} 
 \,\, \sum_{q', \bar q' = -b}^b\,\, 
\sum_{q'',\bar q'' = -b}^b\,\, \sum_{q = Q-b}^b 
  |q',q''\rangle \langle \bar q',\bar q'' | Q- q, q \rangle \langle Q-q,q | \nn \\
  && \qquad \times \exp\left\{ \frac{2\pi \ri}{2b+1}\left(q'P+ q'p + q''p - \bar q'P
  -\bar q' p - \bar q'' p\right) \right\} \nn \\
  &=& \frac{1}{2b+1} \sum_{p = -b}^{b-P}\,\, \sum_{q', q'' = -b}^b 
 \,\, \sum_{q = Q-b}^b |q',q''\rangle \langle Q-q,q | \exp\left\{ \frac{2\pi \ri}{2b+1} 
  \left(q'P + q'p + q''p - QP +qP -Qp \right) \right\}.
\end{eqnarray}
Applying this operator on the initial state $| \Gamma_0 \rangle$ finally yields
\begin{eqnarray}
  \hat \Pi_P \hat \Pi_Q | \Gamma_0 \rangle &=& \frac{1}{(2b+1)^{3/2}} 
  \sum_{p = -b}^{b-P} \sum_{q', q'' = -b}^b \sum_{q = Q-b}^b
  |q',q''\rangle_A \otimes | -q \rangle_B \nn \\
  && \qquad \times \alpha_{Q-q} \exp\left\{ \frac{2 \pi \ri}{2b+1} \left(q'P 
  + q'p + q''p - QP +qP -Qp \right) \right\}\\
&=& \Biggl[\frac{1}{(2b+1)^{3/2}}  \sum_{p = -b}^{b-P} \sum_{q', q'' = -b}^b
\exp\left\{ \frac{2 \pi \ri}{2b+1} \left(q'P 
  + (q'+ q'')p - Q(P-p) \right) \right\}   |q',q''\rangle_A\Biggr]
\otimes \bigl\vert\Gamma_2\bigr\rangle_B
\end{eqnarray}
where $\bigl\vert\Gamma_2\bigr\rangle_B$ is Bob`s 
(unnormalized) state after the measurement of $\hat Q_+$
and $\hat P_-$
\be
  | \Gamma_2 \rangle_B =\sum_{q = Q-b}^b \alpha_{Q-q}\,  
  \exp\left\{{\frac{2\pi \ri qP}{2b+1}}\right\}\,   | -q \rangle_B.
  \label{eqn-Gamma_pre}
\ee
This represents the initial unknown state with 
shifted amplitudes $\alpha_{Q-q} $ and some phase 
factors $\re^{\frac{2\pi \ri qP}{2b+1}}$. Both, the index shift $Q$ 
and the phase factor proportional to $P$ are
known to Alice after the joint measurement of $\hat Q_+$ and $\hat P_-$ 
and the corresponding information can be transmitted to Bob by
classical channels. Bob can then 
apply appropriate shift and phase-rotation operations
to his quantum state to recover a replica of the input state. 
We do see however that due to the index shift and the
finite dimension of the Hilbert space of Bob`s state, some
of the original state amplitudes $\alpha_q$ may be lost, depending on
the value of $Q$. Consequently the
teleportation has only a finite success probability.


\section{Probability of success}


We now discuss the probability of success of the described 
teleportation protocol.
For this we generalize the above 
discussion and drop the requirements $Q \ge 0$ and $P \ge 0$. Then eqn.
(\ref{eqn-Gamma_pre}) can be rewritten as
\be
  | \Gamma_2 \rangle \sim \sum_{q={\rm max}[-b,-b+Q]}^{{\rm min}[b,b+Q]}
\alpha_{q}\,  \exp\left\{{\frac{2 \pi \ri}{2b+1}(q+Q)P}\right\}
\, |q+Q\rangle. \label{Gamma-dis}
\ee
We see that only state amplitudes $\alpha_q$ survive for which 
$q\in\{{\rm max}[-b,-b+Q],{\rm min}[b,b+Q]\}$.
Thus a sufficient condition for a successful teleportation of 
an arbitrary qudit state is that the measured eigenvalue $Q$
fulfills
\be
|Q| \le b-a. 
  \label{condition_success}
\ee
If this inequality holds, the knowledge of $Q$ and $P$ is sufficient for
Bob to reproduce the initial state by 
local operations. 

We now want to calculate the success rate for the given
teleportation protocol, i.e.
the probability that condition
(\ref{condition_success}) is fulfilled. 
The probability to measure a specific eigenvalue $Q$ is
\begin{eqnarray}
  p(Q) &=& \abs{\left\langle Q \right| \left. \Gamma \right\rangle}^2 \nn \\
  &=& \frac{1}{2b+1} \sum_{q_1=-a}^a \sum_{q_2=-b}^b \abs{\alpha_{q_1}}^2 
  \delta_{q_1+q_2,Q} 
\end{eqnarray}
The probability of success of the teleportation protocol is therefore
\begin{equation}
  P =\sum_{Q = -(b-a)}^{b-a}p(Q)=
 \frac{1}{2b+1} \sum_{Q = -(b-a)}^{b-a} \sum_{q_1,q_2}
  \abs{\alpha_{q_1}}^2 \delta_{q_1+q_2,Q}.
\end{equation}
Independent of the value of $\alpha_{q_1}$ the summation over $q_2$ and $Q$
yields
\[
  \sum_{Q=-(b-a)}^{b-a} \sum_{q_2 = -b}^b \delta_{q_2,Q-q_1} = 2(b-a)+1.
\]
This is because $a-b \le Q \le
b-a$ and $-a \le q_1 \le a $ always imply $-b \le Q - q_1 \le b$ and
so one has $2(b-a)+1$ non-vanishing contributions regardless of $q_1$.
With this we eventually arrive at the success probability
\begin{eqnarray}
  P &=& \frac{ 2(b-a)+1 }{2b+1} \sum_{q_1} \abs{\alpha_{q_1}}^2 \nonumber \\
    &=& \frac{ 2(b-a)+1 }{2b+1}= 1 - \frac{2a}{2b+1},
\end{eqnarray}
which does not depend on the input state. 
We can rewrite this result noting that the 
dimension of the Hilbert space of the input state is 
dim$\{{\cal H}_i\}=2a+1$ and the dimension  of the
ancilla Hilbert space is dim$\{{\cal H}_a\}=2b+1$:
\be
  P = 1 - \frac{{\rm dim}\{{\cal H}_i\}-1}{{\rm dim}\{{\cal H}_a\}}.
  \label{eqn-success-dim1}
\ee
Eq.(\ref{eqn-success-dim1}) is the main result of our paper. It shows that a
linear teleportation protocol is in general always probabilistic and that
its success probability only depends on the relative Hilbert dimensions
of the input and ancilla spaces. In the case of qubits, i.e. for
${\rm dim}\{{\cal H}_i\}={\rm dim}\{{\cal H}_a\}=2$ we obtain $P=1/2$ 
as realized
e.g. in the scheme of Weinfurter \cite{Weinfurter-EPL-1994}. Likewise 
in the case of large Hilbert spaces with
${\rm dim}\{{\cal H}_a\}\gg{\rm dim}\{{\cal H}_i \}$ as in the 
Braunstein-Kimble proposal \cite{Braunstein-PRL-1998} in the limit of
perfect squeezing, a unit success 
probability can be achieved.

It should be noted 
that an analogous calculation can be performed in the case of continuous variables
with a constant density of states but bounded spectrum. 
In this case the input and the ancilla 
states are given by
\be
  |\Psi\rangle &=& \int_{-a}^a \!\! {\rm d} q_1 \alpha(q_1) |q_1 \rangle \\
  |\Phi\rangle &=& \frac{1}{\sqrt{2b}} \int_{-b}^b \!\! {\rm d} q_2 \int_{-b}^b \!\!
  | q_2 \rangle | -q_2 \rangle.
\ee
Then $\hat q$ and $\hat p$ are position and momentum operators or 
quadrature amplitudes. One obtains the similar result
\begin{equation}
  P_{\text{cont}} = 1 - \frac{a}{b}.
\end{equation}
The only difference in this formula is the missing of the term $+1$ 
which arises from the different treatment of the end
points.


\section{Efficiency of the KLM teleportation scheme}


At first sight the KLM \cite{Knill-Nature-2001} teleportation scheme seems 
to implement the generalized discrete teleportation scheme discussed 
in this paper where the quantum number $q$ is the total number of
photons in Alice's modes since its probability of success scales as
\be
  P = 1 - \frac{1}{n+1}.
\ee
The entangled ancilla state consist of $n+1$ terms containing 
$n$ photons each. This is exactly the result
one would obtain from eqn. (\ref{eqn-success-dim1})
if we insert dim$\{ {\cal H}_i\} = 2$ (qubits) and 
dim$\{ {\cal H}_a \} = n+1$. 
But this interpretation is not correct because the dimension of the 
Hilbert space ${\cal H}_a$ of the $n$ ancilla photons is much larger. 
Although the true Hilbert-space dimension of $n$ photons distributed over
$2n$ modes is much larger that $n+1$, one could 
argue that in fact only a small subspace of this large Hilbert space is 
actually used. This leads us to the question what is the relevant dimension of 
the Hilbert space of the ancilla photons. 

One definition that gives a lower bound for the dimension of the used
Hilbert space is the number of distinguishable measurement outcomes.
In the KLM case one applies a discrete $n+1$ point Fourier transformation
on the first $n+1$ modes followed by photodetection of these modes. One
has to measure not only the total photon number but also their distribution 
over the modes. In fact every possible 
distribution of $k = 0,\ldots,n+1$ 
photons over $n+1$ modes does occur. As photons in the same mode are 
indistinguishable the number of different measurements is smaller 
than the physical dimension of the Hilbert space and one has 
a number of
\be
  N_{\rm KLM} &=& \sum_{k = 0}^{n+1} \left({n+k} \atop {k} \right) \nn \\
  &=& \frac{(2n+2)!}{((n+1)!)^2}
\ee
different measurement outcomes. In contrast, the measurement of $\hat Q_+$
and $\hat P_-$ discussed in this paper can only result into $2(a+b)+1$ 
resp. $4b+1$ different measurement outcomes. I.e. one can distinguish 
\be
  N = \bigl( 2(a+b)+1 \bigr) \left(4b+1 \right)
\ee
different measurement results. We see that for the discussed teleportation
scheme the number of distinguishable measurement outcomes scales quadratically 
in $b$ whereas for the KLM teleportation scheme this number scales much less
favorable with
$(2n+2)!/((n+1)!)^2$.
Since the KLM scheme does not 
scale optimal in the sense of the probability of 
success calculated above, 
it seems feasible that other linear teleportation
schemes may be developed 
whose success probability scales better.
This is of particular interest for the practical implementation of linear 
optical quantum computing, for example in the protocol described by Yoran 
and Reznik \cite{Yoran2003}, where the resources depend critical on the 
scaling of the probability of success of the single quantum gates.


\section{Success vs. Fidelity}


Recently Franson, {\it et al.}~\cite{Franson-PRL-2002} suggested a modification 
of the KLM scheme which always succeeds but whose fidelity, i.e. the
overlap of the teleported to the input state is not unity. According to
reference \cite{Franson-PRL-2002} we will rather consider the square 
of the fidelity
\be
  F = \left| \left\langle \Psi_{\rm in} \right. \left| \Psi_{\rm out}
   \right\rangle \right|^2.
\ee
In the proposal of Franson, {\it et.al} the square of the fidelity 
scales as $F=1-1/n^2$. This approach is different from the
present one and the one of Knill, Laflame, and Milburn, where the teleportation
reproduces exactly the input state if it is successful. Whether or not the
teleportation is successful is hereby uniquely determined by the measurement
result. This has the advantage that 
by post-measurement selection of a teleported ensemble a sub-ensemble 
of exact replica of the input state can be generated. The obvious advantage
of the Franson scheme is that in any case a state is teleported which 
is similar to the input one. 

To make a comparison to the Franson scheme we could ask the question
what is the average fidelity of our scheme if we do not discard 
those events where $Q$  
falls outside of the 
interval given by eq.(\ref{condition_success}). 
Then the mean 
squared fidelity of the discussed teleportation scheme is:
\begin{eqnarray}
  \bar F &=& \sum_{Q = -(b+a)}^{b+a} p(Q) \bigl\vert\langle 
            \Gamma_2^Q | \Psi \rangle\bigr\vert^2 \nn \\
         &=&  P \cdot 1 + \sum_{\abs{Q} > b-a} p(Q)\bigl\vert
\langle \Gamma_2^Q 
            | \Psi \rangle\bigr\vert^2       
\end{eqnarray}
where $| \Gamma_2^Q \rangle$ is Bob's state after the measurement 
with result $Q$. $p(Q)$ denotes the probability 
to obtain this outcome and $P$  denotes the probability of 
success derived in 
previous sections, corresponding to an exact reproduction of the
input state. 
To estimate the fidelity we assume in the following that
the basis states $| q_i \rangle $ have an equal a priori 
probability to appear in the input state of the teleportation 
such that we have the mean value of the coefficients $\alpha_{q_i}$:
\be 
  \langle \abs{\alpha_{q_i}} \rangle = {\rm const.} = \frac{1}{\sqrt{2a+1}}
\ee
Then given a measurement outcome with $Q > b-a$,
the overlap between the input and the output state is
\begin{eqnarray}
 \bigl\vert \langle \Gamma_2^Q | \Psi \rangle\bigr\vert &=& 
  \frac{1}{2a+1} \sum_{q = Q - b}^{b} \sum_{q' = -a}^a \delta_{q,q'} \nn \\
  &=& \frac{a+1+b-Q}{2a+1}.
\end{eqnarray}
Considering as an example qubits that can only assume two values (i.e. 
$a = 1/2$) 
the teleportation does not reproduce the exact input state iff $Q
= b + 1/2$ or $Q = -b - 1/2$. In this case it is $\bigl\vert
\langle \Gamma_2^Q | 
\Psi \rangle\bigr\vert^2 = \frac{1}{4}$ and $p(b+1/2) = p (-b-1/2) = 
\frac{1}{2(2b+1)}$ and thus we find for the mean squared fidelity 
of our teleportation protocol:
\be
   \bar F &=& 1 - \frac{1}{2b+1} + \frac{1}{4(2b+1)}.
\ee
The first two terms represent the non unity  probability 
to exactly reproduce the input state, whereas the last term
describes the finite but nonvanishing overlap of the teleported
state with the initial state in the previously unsuccessful cases.
We see that allowing for a nonperfect reproduction of the state
the fidelity can be enhanced as compared to the result of the last section.
Still $1-F$ scales linear in $1/b$ as compared to the quadratic scaling
in $1/n$ of the Franson scheme. 
One has to keep in mind however that the number $n$ in the KLM
or Franson scheme is not the dimension of the ancilla Hilbert space,
as we pointed out in the preceding section.


\section{Conclusion}


In the present paper we discussed the question why continuous 
teleportation can be performed with linear elements in a determinstic way
while the discrete counterpart requires nonlinear elements in form 
of quantum gates in order to be successful in all cases.
For this we introduced a teleportation protocol
similar to that used in the continuous case applied to qudits with
a discrete and finite set of basis states. We also generalized the
notion of linear elements and detection to the measurements of operators
$\hat Q_+=\hat q_1+\hat q_2$ and $\hat P_- = \hat p_1-\hat p_2$ which are 
linear combinations of the basic observables $\hat q$ and $\hat p$ in the
input and ancilla spaces. We have shown that this protocol which uses
only linear elements allows only for a probabilistic teleportation, a result
recently shown to be true for any linear protocol \cite{Luetkenhaus-preprint},
and has a success probability $P$ which is determined only by the Hilbert space
dimensions of the input (${\rm dim}\{{\cal H}_i\}$) and ancilla states 
(${\rm dim}\{{\cal H}_a\}$). In the case of qubits we recover
the value $P=1/2$, which is the known limit for discrete teleportation with
linear elements. On the other hand for the continuous teleportation
protocol of Braunstein and Kimble the requirement for a local oscillator
field with infinite squeezing, and thus infinite photon number, implies that 
the effective
dimension of the ancilla Hilbert space is much larger than that of the
input space. In this case the success probability of linear teleportation
approaches unity. Thus the difference between discrete and
continuous teleportation appears to result from the difference
in the ancilla resources used rather than hidden nonlinearities 
as suggested in \cite{Vaidman-PRA-1999}.

We have also compared the Knill Laflame Milburn
proposal for linear teleportation 
\cite{Knill-Nature-2001}
as well as the one by Franson and coworkers
\cite{Franson-PRL-2002}
with our abstract scheme and found that both do not scale optimal. 
This suggests that it may be possible to construct specific linear
teleportation schemes for photons which use much less resources than the
KLM and Franson schemes. 


\section*{Acknowledgment}


We would like to thank Norbert L\"utkenhaus and Peter van Loock
for stimulating discussions and making results available prior to 
publication. The financial support of the Deutsche Forschungsgemeinschaft
within the Schwerpunktprogramm ``Quanteninformation'' is gratefully
acknowledged.






\begin{thebibliography}{99}


\bibitem{Knill-Nature-2001} E. Knill, R. Laflamme, G.J. Milburn, Nature {\bf 409}, 
46 (2001).

\bibitem{Bennett-PRL-1993} C.H. Bennett {\it et al.}, Phys. Rev. Lett.
{\bf 70}, 1895 (1993).

\bibitem{Lutkenhaus-PRA-1999}    
N. L\"utkenhaus, J. Calsamiglia, and K.-A. Suominen, 
Phys. Rev. A {\bf 59}, 3295 (1999). 

\bibitem{Calsamiglia-PRA-2002} J. Calsamiglia, Phys. Rev. A {\bf 65}, 030301 (2002)

\bibitem{Cerf-PRA-1998} N.J. Cerf, C. Adami, and P.G. Kwiat, Phys. Rev. A {\bf 57}
R1477 (1998)

\bibitem{Kwiat-PRA-1998} P.G. Kwiat, H. Weinfurter, Phys. Rev. A {\bf 58}, R2623 (1998).

\bibitem{Weinfurter-EPL-1994} H. Weinfurter, Europhys. Lett. {\bf 25}, 559 (1994).

\bibitem{Braunstein-PRL-1998} S.L. Braunstein and H.J. Kimble, Phys. Rev. Let. 
{\bf 80}, 869 (1998).

\bibitem{Vaidman-PRA-1999} L. Vaidman and N. Yordan, Phys. Rev. {\bf 59}, 116 (1999).


\bibitem{Vaidman-PRA-1994} L. Vaidman, Phys. Rev. A {\bf 49}, 1473 (1994).


\bibitem{Franson-PRL-2002} J.D. Franson {\it et. al.}, Phys. Rev. Lett. {\bf 89},
137901 (2002).



\bibitem{Werner-JPA-2001} R.F. Werner, J. Phys. A {\bf 34}, 7081 (2001).

\bibitem{Yoran2003} N. Yoran and B. Reznik, quant-ph/0303008 (2003).

\bibitem{Luetkenhaus-preprint} Peter van Loock and Norbert L\"utkenhaus,
quant-ph/0304057 (2003).



\end{thebibliography}
\end{document}